# Heavy Long-lived Mössbauer State of Niobium

[1,2]Yao Cheng (程曜) and [3,4]Chi-Hao Lee (李志浩)

[1*]Department of Engineering Physics, Tsinghua University, Beijing, Haidian, 100084, China

[2]Center for Micro and Nano Mechanics, Tsinghua University, Beijing, Haidian, 100084, China

[3]Department of Engineering and System Science, National Tsinghua University, Hsinchu, 300, Taiwan

[4]National Synchrotron Radiation Research Center, Hsinchu, 300, Taiwan

**Abstract**

A heavy niobium state showing 1/3 residual resistance is discovered below the superconducting transition temperature. This niobium sample contains high-density long-lived Mössbauer excitation.

Previous studies on $^{103m}$Rh have shown many unusual findings, which reveal a collective excitation of the long-lived Mössbauer state [1-6]. The reported findings include: Rabi frequency as a function of temperature and $^{103m}$Rh density; anomalous emissions with wide-spreading and peaking structures; anisotropic nuclear susceptibility depending on the sample geometry at room temperature; and a quantum phase transition showing collapse and revival. The underlying physics is the nuclear resonance provided by the simultaneous emission of two entangled photons, called biphoton. Rhodium has only one isotope, which provides a nature-made photonic crystal. Biphoton wave vectors are free to match lattice constants, which enables the collective coupling without recoil. Strong collective coupling (>50 eV) between biphoton and nuclei disables the scattering with phonon and electron. In other words, the long-lived Mössbauer resonance is thus possible because of multi-photon transitions in a crystal consisting of identical nuclei. Silver is obviously not the best case [6] because silver has two kinds of natural isotopes. Biphoton confined in the photonic crystal creates a mass of eV order calculated using the critical density of the Bose-Einstein condensation (BEC) [3]. Due to this light mass, the so-called nuclear spin-density wave (NSDW) exhibits a magneton six orders of magnitude bigger than the Bohr magneton [3]. NSDW carries no charge but spin, which provides a texture dictated by the macroscopic sample geometry [2,3]. The most likely ordering of NSDW is anti-ferromagnetic because it is a boson.

Resolving a long-standing problem of the long-lived Mössbauer effect with varied interpretations across disciplines leaving physicists unconvinced has motivated us to study $^{93m}$Nb, as an alternative long-lived Mössbauer candidate. Niobium and rhodium have several features in common such as both of them are single isotopes containing the Mössbauer state. The rhodium Mössbauer emission has energy of 39.8 keV, a half-life of 56 minutes, and an E3 transition, while niobium Mössbauer emission has energy of 30.8 keV, a half-life of 16 years, and M4 + E5 transitions [7]. We calculate the Rabi frequency of $^{93m}$Nb using the lifetime, the internal conversion rate [7], and the E5 transition (detailed later) from the reported Rabi frequency of $^{103m}$Rh [5]. This simple estimation gives a vacuum Rabi frequency of $^{93m}$Nb of the order of 10 μeV. A 10-μeV strong coupling is unable to survive from electron scattering. However, we anticipate the nonlinear increase of the coupling strength with the $^{93m}$Nb density [5]. High-density NSDW survives when its Rabi frequency is faster than the phonon decay time [2] and the scattering time with electrons. NSDW and electronic Cooper pair interact because both are entangled with the longitudinal phonon. Obviously, if NSDW were formed with low-density $^{93m}$Nb, its impact on

---

electronic Cooper pairing would have already been observed in the superconducting cavity of accelerators [8]. Sufficient NSDW density is thus required to have new physics.

We prepared a single-crystal Nb sample (oval plate of 1.2mm×12mm×13mm) by neutron irradiation with $10^{12}$ n/cm$^2$s for 5 hours in the reactor at Tsinghua University, Hsinchu in 2008. The resulting counting rate of K-shell x-rays was 20 counts per second into the 4π solid angle in 2008. At the time of experiments, the $^{93m}$Nb density was $2\times10^{13}$ cm$^{-3}$. We are unable to give precise $^{93m}$Nb density due mainly to the roughly estimated ratio of internal conversion in ref. [7]. We used a polycrystalline sample without irradiation as reference. The irradiated sample was annealed together with the reference sample at 1400 °C for two hours to re-grow the crystal grains [8]. We measured the bulk resistance by means of a standard four-point method using ac driving current. The signal was delayed due to a large heat capacitance. We minimized this issue by soldering an indium thermal contact between the samples and the cold head. To avoid two grounds on the cold head and the amplifiers, we inserted two 10:1 transformers to isolate the grounds. The samples were mounted with their normal vectors in the E-W direction on the cold head. The current of 50 mA and the voltage pick-ups were horizontally applied in the N-S direction.

Figure 1 compares the resistance of the irradiated single crystal and the reference. We observed a slightly reduced transition temperature for both samples. Control against dc-measurements of the small pieces cut from the samples indicates that the temperature shift is due to the signal delay, as mentioned above. With the 6-Hz driving current, the resistance shows a constant plateau with 1/3 residual resistance from 7.7 to 8.5 K. As shown in the inset of Fig. 1, the plateau contains both a real part and an imaginary part, which reveals that the voltage phase is slightly ahead of the current phase. This plateau is verified by comparing curves of the cooling and warming periods in Fig. 2. During the cooling period, there was a significant temperature variation ΔT ~ 1 K with the pumping frequency of 1-Hz at the cold head. This temperature variation created the zigzag plateaus of the real and the imaginary parts from 9 to 9.5 K, as shown in Fig. 2. Cooling and warming curves both demonstrate the plateau, indicating that they are not artifacts. The real curves coincide at 9 K in Fig. 2, revealing the proper reading of the temperature sensor on the cold head.

We conclude from this the existence of a new state, which is sensitive to frequency and temperature.

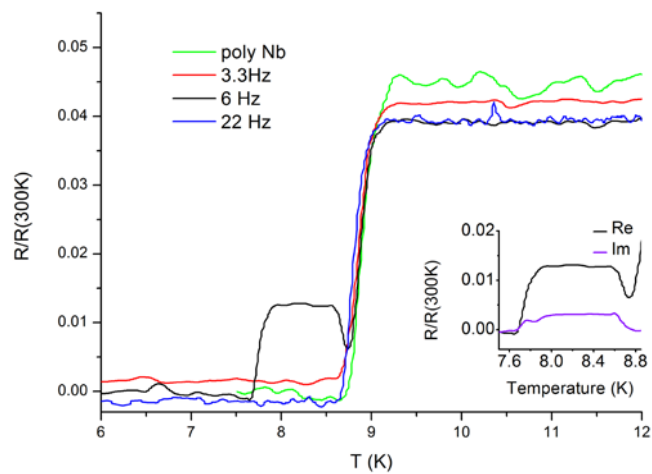

FIG 1: Resistance normalized by the room-temperature R(300K) as a function of temperature. Four measurements are shown: 1, reference sample with 3.3 Hz (green); 2, irradiated sample with 3.3 Hz (red); 3, irradiated sample with 6 Hz (black); 4, irradiated sample with 22 Hz (blue). The inset gives the real (black) and imaginary parts (violet) of the 6-Hz plateau. To guide the eyes, 22-Hz curve is shifted downwards and 3.3 Hz curve is shifted upwards. The ordinate scale stands for the black and green curves sitting on the baseline.

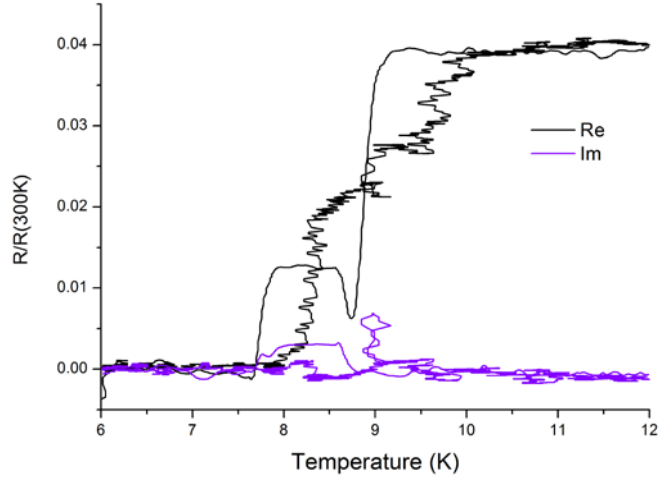

FIG 2: Real (black) and imaginary (violet) parts of cooling (zigzag line) and warming periods (straight line) of the 6-Hz measurement.

Biphoton has an intrinsic spin of 2 and an even parity. To conserve the parity, $^{93m}$Nb emits a biphoton via the E5 transition. The quantum numbers of NSDW are J=4, L=3, and S=2. Because of the $^{93m}$Nb transition between 1/2- state and 9/2+ state [7], NSDW carries 4 spin quanta with an odd parity. The horizontally applied current has an angle of 40 degrees to the 50-μT earth field $B_E$ in Beijing. NSDW under magnetic field split into 9 Zeeman levels. So long as NSDW has a magnetic ordering, the spin-spin interaction provides itinerary electrons a periodic potential. Electrons prefer triplet pairing in the presence of NSDW, which is then split into 3×9 Zeeman levels. The occupation number of NSDW in each Zeeman level is volume dependent, and we estimate it to be $10^{11}$ in the bulk sample. The longitudinal phonon is absorbed by the longitudinal mode of NSDW that provides a mass ~2 eV for $6\times10^{11}$ cm$^{-3}$ per spin multiplicity, estimated from the BEC density reported previously [3], where the mass is assumed to be inversely proportional to the inter-particle spacing. In the absence of triplet pairing of electrons at room temperature, the NSDW system already undergoes BEC.

The plateau below the transition temperature in Fig. 1 is due to the interaction between NSDW and pairing electrons, likely via the longitudinal phonon. This gives us further information about the mass of NSDW and the scattering of pairing electrons. Two thirds of the current is free from scattering. One third of the current carried by the interacting pair is responsible the plateau height, of which the imaginary part indicates a resonant frequency ω slightly less than 6 Hz. The mass m of g×64 GeV is calculated from the Zeeman splitting $\omega=egB_E/2m$. The g-factor of massive biphoton shall be > 1 due to its internal structure, as revealed by proton and neutron. The big mass number > 64 GeV may indicate flipping of the whole NSDW condensate consisting of two identical particles in circulation.

Note:

We observed a significant fluctuation of voltage atop the baseline, when the irradiated sample was kept at 4 K and measured with 1.1 Hz. This voltage fluctuation was measured for half an hour, showing a frequency spectrum with one peak at 0.006 Hz. A peak with the same frequency also showed up several times at the fluctuation of the Rabi frequency, when the rhodium sample reached room-temperature BEC and cooled down to 77 K [1-5].

Acknowledgements:
We thank Cheng-Chung Chi (齐正中), Jia-Rui Yang (杨佳瑞), Danier Skieller, Ping Xue (薛平), Ru-Fan Zhang (张如范) and Fei Wei (魏飞). We also thank Hai-Hu Wen (闻海虎) and Bing Shen (沈冰), institute of physics CAS, for the dc measurement; Hua-feng Wang (汪华峰) and Kai-You

Wang (王开友), Institute of semiconductors CAS, for the cooling experiments; Zhong-Quan Li (李中泉), Institute of high energy physics CAS, for the experiment with liquid helium. We thank Josef Ellingsen for polishing the English writing.